\documentstyle[aps,prl,multicol]{revtex}
\begin{document}
\draft
\title
{\Large\bf A Kondo impurity 
in one dimensional
correlated conduction electrons}
\author{Xiaoqun Wang}
\address{Max-Planck-Institut f\"ur Physik komplexer Systeme,
Bayreuther Str. 40 Haus 16 D-01187 Dresden Germany}
\date{\today}
\maketitle
\begin{abstract}
A spin-$\frac 1 2$ magnetic
impurity coupled to a one-dimensional correlated electron system have been studied
by applying the density renormalization group method.
The Kondo temperature is substantially enhanced by 
strong repulsive interactions in the chain,
 but changes non-monotonically in the case of electron attraction.
The magnetization of the 
impurity at zero-temperature shows local Fermi
liquid behavior.

\end{abstract}
 \begin{multicols}{2}
During the last decade, much effort has been devoted to the
investigation of the electron-correlation effects. They are
present most prominently in the 
copper-oxide materials which display
high-$T_c$ superconducting and unusual normal-state properties\cite{R1}
 and may even show  heavy-fermion behavior\cite{R2}.
In one dimension (1D), 
an interacting electron system is known to be 
a Luttinger liquid \cite{R3}. 
It seems possible that in the future by means of 
quantum wire fabricated
by nanotechnique one can study the effect of magnetic impurities
on Luttinger liquid\cite{R4} . 

The problem of a Kondo 
impurity coupled to a Luttinger liquid was first considered 
by Lee and Toner\cite{R5}.
By using perturbative renormalization group
theory they found that the Kondo temperature $T_K$ 
depends on the coupling $J$ between the impurity and the
conduction electrons in the form of a power law (not
exponential) provided that the interaction between electrons 
is sufficiently strong compared
to $J$. Because of the neglect of local backscattering 
their treatment does not preserve SU(2) symmetry.
The local backscattering was included within the poor man's scaling by
Furusaki and Nagaosa \cite{R6} who obtained
a stable strong coupling fixed point for $J>0$ as well as $J<0$.
Moreover, by employing a $1/J$ expansion, they found
that a temperature expansion of thermodynamic quantities shows
critical behavior, i.e., {\it non-integer exponents}. It is an open
question though, how well such a perturbative expansion
 is justified in the strong coupling regime. In fact, a
simplified model\cite{R7} for conduction electrons, which consists
of right-moving spin-up and
left-moving spin-down electrons, leads to the conclusion that the Kondo
impurity behaves like a local Fermi liquid\cite{Noz}. 
On the other hand,  boundary conformal field theory methods\cite{R8,R81} result in
two types of critical behavior, i.e., 
either a local Fermi liquid or one of the form proposed in \cite{R6}.
Most recently, Chen et al\cite{R9} predicted local Fermi liquid
 behavior by making use of 
the parity and spin-rotation symmetry of the problem.

The above literature indicates that the critical 
behavior of a Kondo impurity
coupled to a Luttinger liquid 
is not fully understood yet. One scenario would be that of a local Fermi
liquid
but with a substantially enhanced $T_K$ due to strong 
correlations between conduction electrons. For example, 
the characteristic energy scale in
the heavy fermion system Nd$_{2-x}$Ce$_x$CuO$_4$ is much larger
than expected for Kondo ions coupled to
free electrons\cite{RR2}. 

In this paper, 
we study ground-state properties of
a Kondo ion antiferromagnetically coupled to 
a 1D Hubbard model by applying the Density Matrix
Renormalization Group (DMRG)\cite{R10}.
The Hamiltonian is:
\[ \hat H_0=-t\sum_{i,\sigma}[\hat c^{\dagger}_{i\sigma}
\hat c_{i-1,\sigma}
+{\rm h.c}]+U\sum_{i}\hat n_{i\uparrow}\hat n_{i\downarrow}+
J\hat{\bf S}\cdot\hat {\bf s}_0\]
where $\hat {\bf S}$ is the impurity spin, $\hat{\bf s}_0$ 
an electron spin at $i=0$ and other notations are standard.
In the limit of $U=\infty$ and half-filling, 
the problem reduces to one of an impurity coupled to a Heisenberg
antiferromagnetic chain, a system studied before (see Refs. 
\cite{RJK,RIG,EA}). Here we will consider all $U$ values and band
fillings equal to as well as different from one-half. 

The Kondo problem is concerned with the 
question of how the magnetic moment of the impurity is suppressed by
decreasing temperature $T$ or external magnetic field $H$.
The critical behavior of a magnetic impurity shows up 
in the low-$T$ thermodynamics
{\it and also } low-$H$ properties\cite{R14,R13,TW,Hew}.
Here we would like to determine the latter for $T=0$.
For that purpose, we calculate the spin 
susceptibility $\chi_0$ of the impurity
at $H=0$ and 
the magnetization of the impurity for various values of $H$. The
former gives rise to the energy scale 
(the Kondo temperature) $T_K$ 
or alternatively the screening length $\xi_K$ ($\chi_0\propto
1/T_K\propto\xi_K$).
The field-dependent magnetization $M(H)$ exhibits
the critical behavior of the system, since it contains a term
proportional to $(H/T_K)^{\alpha}$ with $2<\alpha\le 3$ for small $H$. 
This can be used to distinguish between a local Fermi liquid \cite{R7,R9}
or an anomalous response case\cite{R6}.  As the field is applied 
exclusively to the impurity spin, we add $ \hat H_M=-H\hat S_z$ 
to $\hat H_0$. By using the Hellmann-Feynman
theorem, we have  
 $M(H)=\left<\Psi_0(H)\right|\hat S_z\left|\Psi_0(H)\right>$
where $\left|\Psi_0(H)\right>$ is the ground state in the
presence of $\hat H_M$, and $\chi_0=\left[\partial M(H)/\partial
  H\right]_{H=0}$. $\chi_0$ is numerically evaluated as the coefficient
of the linear term in an $H$-expansion of $M(H)$.
For given $J$ and $U$, the values of $H$ is chosen so small, 
for instance $H=0.00015$ at $J=0.5$ and $U=20$, that the 
contributions from higher orders in $H$ are negligible
in comparison with numerical errors (see below). 

The application of the DMRG method 
to inhomogeneous systems as considered here is not
straightforward 
because local couplings must be properly
renormalized by DMRG procedures\cite{R11}.
Let us discuss briefly some
essential points involved in our studies.
The number of states kept mostly is between 256 and 512.
The truncation errors are the order of $10^{-10}$. 
In order to work with a non-degenerate ground state, 
our system consists of one impurity and 
an odd number of conduction electrons $N$. 
The filling factor $\nu=N/L$ ($L$: the number of sites) equals one, for which 
DMRG calculations are performed with odd $L$ under
open boundary conditions, if not
stated otherwise, but values of $\nu=1/2$ and $3/4$ and correspondingly
an even number of sites $L$ are also
considered with periodic boundary conditions. 
When $L=4k+1$ ($k$ being an integer) and the impurity
is located to the centre of the chain the ground state has total spin 
$S^{tot}=0$\cite{Spin}. 
In this case, the two sites which are added in each
DMRG step have maximum distance to the impurity,
and at least three {\it sweeps} are taken when the
calculations are performed. 
 In the presence of $\hat H_M$, the ground state is
again taken in the $S^{tot}_z=0$ subspace. 
Systematic relative-errors based on truncations 
are estimated to be the order of $10^{-7}$. 
DMRG calculations are done for chain lengths $33\leq L\leq105$ so
that independent extrapolations show relative deviations between
$10^{-4}$ and $10^{-3}$. We consider those as 
the systematic relative-errors
 of our final results.

The spin of a Kondo ion is compensated by a spin screening cloud
surrounding the ion\cite{R12}. When $U=0$, the 1D system is
metallic. The screening length $\xi_K$ is then
related to the Kondo temperature $T_K$ via $\xi_K\approx v_F/T_K$
 where $v_F$ is the
Fermi velocity. When $J\rho\ll W$
where $W$ is the  band width and $\rho=1/2\pi$ is the density of state 
at the Fermi level ($t=1$ in our calculation), 
it is $\xi_K\approx a e^{1/J\rho }/\sqrt{J\rho}$.
For $T_K\sim 10K$,
 the screening length extends over thousand lattice spacings $a$.
The strong correlations between conduction electrons decrease $\xi_K$
{\it substantially}. 
Figs. 1(a), and (b) show $\chi_0\sim \xi_K(J)$ [also $\chi_0^{-1}\sim T_K(J)$] 
for $U=0$, and $20$, respectively. i) {\it a crossover:} 
Consider first Fig. 1(a). In order to reach the thermodynamic limit,
we must have $L\gg \xi_K/a$.
This limits us to $J\ge2$. The figure suggests a crossover from
a strong coupling to a weak coupling at $J\approx 2.5$. In Fig. 1(b),
the same crossover takes place at approximately $J\approx0.4$. 
ii) {\it a weak coupling regime:} Note that for
$J=0.3$ we find $\chi_0=3.73$ for $U=20$ while for $U=0$ we obtain
$\chi_0\sim 10^5$ \cite{R12}. iii) {\it a strong coupling regime:}
One also notices from Fig. 1(a), and (b) that for large values of $J$ it is 
$T_K\sim J$. The linear dependence sets in for $U=20$ at much lower 
values of $J$ ($J\stackrel{>}{\mbox{\tiny $\sim$}} 1$) than for $U=0$ 
($J\stackrel{>}{\mbox{\tiny $\sim$}} 40$). 
These features on the energy scale $T_K$ 
support previous findings
\cite{R5,R6,R7,R8,R9}. 

In Fig. 2, we  shows $\chi_0$ as a function of $U$ for $J=4$.
For $U\gg W$,
the exchange coupling between sites in the chain $J_{ex}=1/4U$ 
becomes much smaller than $J$, implying
that a singlet 
 which extends over
{\it a few lattice spacings}
 is formed between the spins of 
the magnetic ion and the chain.
$\chi_0$ saturate at $1/2J$ as $U\gg W$, which
corresponds to a local singlet between the spins of 
the ion and the lattice site $0$. 
The situation differs in the case of electron attraction ($U<0$). 
One notices a maximum in $\chi_0$ at $U_c\approx-1.0$.
For $U_c(J)<U<0$ the effective Kondo coupling is reduced by the
attractive electron-electron interaction, and $\xi_K$ is enhanced. 
With increasing attraction
electrons form more and more on-site pairs. When $|U|$ becomes much
larger than the binding energy of the Kondo singlet and than $W$, all the
electrons are paired except one (remember that $N$ is an odd number).
This unpaired electron forms a singlet with the magnetic impurity so
that $\chi_0$ saturate in the limit $U\rightarrow -\infty$ \cite{R188}. 
Moreover, close to $U=0$, the susceptibility $\chi_0$ is {\it linear in $U$},
which confirms perturbation results for $T_K$\cite{KF}.
When both $J\rho$ and $|U|$ are much smaller than $W$,
one can approximately 
transform the problem into one of an impurity coupled to free electron
with an effective Kondo coupling given by 
$J^z_{eff}=J+U/4$ and $J^{\perp}_{eff}=J$. 
It is i) $J^z_{eff} >J^{\perp}_{eff}>0$ when $U>0$\cite{R9};
ii) $0<J^z_{eff} <J^{\perp}_{eff}$ when $U_c<U<0$; 
iii) $J^z_{eff}<0$ and $J^{\perp}_{eff}>0$ when $U<U_c$. This
classification explains qualitatively the behavior of $\chi_0$ in Fig.
2 in the relevant range of $U$. To explore the behavior of $\chi_0$ 
around $U_c$, we have calculated several values of $U$ near $U_c$. As 
shown in Fig. 2, the curvature close to $U_c$
reveals, with a relative error $\stackrel{<}{\mbox{\tiny $\sim$}}10^{-4}$, 
 that $U_c$ is a crossover rather than a critical point.

It is elucidating to study in more details the different states
between the impurity and the site $0$, which appear in the 
$\left|\Psi_0(H=0)\right>$ as well as their weights.
They are constructed from the impurity-spin states
$\{\left|\uparrow\right>$, $\left|\downarrow\right>\}$
and the spin states $\{\left|0\right>_0$, $\left|\uparrow\right>_0$
$\left|\downarrow\right>_0$, $\left|\uparrow\downarrow\right>_0\}$
of site $0$, from which we can form a 
singlet 
$\left|\psi_s\right>=\frac 1{\sqrt 2}(\left|\uparrow\right>_0
\left|\downarrow\right>-
\left|\downarrow\right>_0\left|\uparrow\right>)$; 
a triplet 
$\left|\psi_t\right>=\{ \left|\uparrow\right>_0
\left|\uparrow\right>$, $\frac 1{\sqrt
2}(\left|\uparrow\right>_0
\left|\downarrow\right>+\left|\downarrow\right>_0
\left|\uparrow\right>)$,
 $\left|\downarrow\right>_0\left|\downarrow\right>\}$; 
and a quadruplet 
$\left|\psi_q\right>=\{ \left|\uparrow\downarrow\right>_0
\left|\uparrow\right>$,
$\left|\uparrow\downarrow\right>_0
\left|\downarrow\right>$, $\left|0\right>_0
\left|\uparrow\right>$,
 $\left|0\right>_0\left|\downarrow\right>\}$.
We extract the these states from 
$\left|\Psi_0(H=0)\right>$ by making use of a
reduced density matrix. The corresponding probabilities are 
$P_s$, $P_t$ (for each of the three states) and $P_q$.
When $U\ll U_c$, one finds $P_s \gg P_t$ and $P_q\gg P_t$, 
and also has $P_s\gg P_q$ if $J$ is a sufficiently large. 
The following results are found when $U\ge0$:
i) for $U \stackrel{<}{\mbox{\tiny $\sim$}} W$, 
one finds
$P_s\ge P_q\ge P_t$; ii) for $U\gg W$, it is 
$P_s\ge P_t\ge P_q$. When in addition, 
$J_{ex}\gg J$, we obtain $P_s\sim P_t\gg P_q$, indicating that
the impurity spin is almost fully polarized by the strong 
correlated electrons in the chain. In contrast, when
$J_{ex}\ll J$, we find $P_s\gg P_t\gg P_q$, implying that a local 
 Kondo singlet is formed.
Two examples are given in Fig.3 for $U=20$ and $J=0.0005$, $0.5$
where also correlation functions
$\left<\Psi_0(H=0)\right|\hat S^z\hat s_i^z\left|\Psi_0(H=0)\right>$ are
shown as a function of $i$. The correlations
are long ranged, particularly for small values of $J$.

The magnetization of the impurity is well suited for studying the critical
behavior of the system under investigation. Shown in Fig. 4 is the
impurity magnetization $M$ as functional $\chi_0H$ ($\sim H/T_K$) for
various values of $J$ and $U$. The solid line corresponds to the
expected behavior in the strong coupling limit, i.e., 
$M(\chi_0 H)=\chi_0H/\sqrt{1+4(\chi_0H)^2}$. 
One notices that the data fall onto a universal curve with slight
 but systematic deviations for $\chi_0H
\stackrel{>}{\mbox{\tiny $\sim$}} 10^{-1}$
(see inset). 
Two different regimes are 
clearly distinguishable,
a strong coupling regime for lower values of $\chi_0H$
and a weak coupling regime in which
the magnetization is saturated at $M=1/2$.\cite{R14,R13,TW}
In the strong coupling regime the impurity spin is well compensated by
the spins of the electrons in the chain, 
while in the weak coupling regime the singlet
is broken by $H \ge \chi_0^{-1}$. 
The deviations are natural even for $U=0$, 
since  
the calculations are performed in real space without any assumption on 
the density of states and the values of $J$ are not much smaller than 
$W$\cite{R199}. 
This however does not change the 
universality class. Particularly,
for small values of $\chi_0H < 10^{-1}$, i.e., the critical 
behavior, $M(\chi_0H)$ behaves 
in the same way for {\it positive and negative} $U$ and 
{\it small and large} $J$ values at  $\nu=1$.
For the case of $\nu\neq 1$, 
let us consider the lower field behavior quantitatively.
According to  \onlinecite{R6},
one might expect that for small fields $H$, $M(\chi_0 H)=\chi_0 H+\alpha_1 
(\chi_0H)^{1/K_{\rho}+1}+{\rm higher~orders~in}~H$. 
When $U$ is finite and $\nu\neq1$, one has $1/2<K_{\rho}<1$.
It turns out that $\chi''(\chi_0H)={\partial}^3M / 
{\partial} (\chi_0 H)^3$ 
should become {\it singular} at $H=0$ if the above
conjecture as regards $M(\chi_0H)$ holds. 
In this case, a scaling analysis is valid.
For finite $L$, the original question becomes
whether $\chi_0''(\chi_0(L)H)\propto
[\chi_0(L)H]^{\alpha(K_{\rho}(L))}$ {\it with or without}
the size-dependent $K_{\rho}(L)$ for the Hubbard model, i.e.,
$\chi_0''(\chi_0(L)H)$ is divergent 
as $\chi_0(L)H\rightarrow 0$
{\it if and only if} $K_{\rho}(L)$ appear in $\alpha(K_{\rho}(L))$.
We have computed 
$\chi''(0)$ for $\nu=1/2$, $3/4$, and $1$ by the exact diagonalization
for the length $L$ up to 12.  
$\chi''(\chi_0H)$ is accurately evaluated 
order by order in a numerical way with the use of 
$\frac {{\rm d}f(x)}{{\rm d}x}\sim (f(x)-f(x-\delta x))/\delta x$ for
 sufficiently small $x=\chi_0(L)H$ and 
$\delta x=\chi_0(L)\delta H$. 
For instance, at $\nu=3/4$ ($L=12$), 
$U=0.5$, and $J=0.01$, $\chi''(x)=-12.000173$, $-12.000157$, $-11.999287$ 
$-11.996831$, and $-11.971354$ for $x=1.476262\times(10^{-5},10^{-4},10^{-3},
2\times10^{-3},6\times10^{-3})$, respectively, 
and $\delta x=6.6020\times10^{-5}$. 
We found that $\chi''(\chi_0H)=-12.00$ at $H=0$ is
{\it independent of} the values of $\nu$, $U$ ($U=0$, $0.5$,
$10$) and $J$ ($J=0.01$, $0.5$, $5$).
For a larger system of $L=33$ and $N=25$,
we obtained $\chi''(\chi_0H=0)=-11.5$ by the DMRG calculation with keeping
800 states and nine sweeps.  
Note that although this result is less accurate than that given by 
the exact diagonlaization, it is {\it still finite}.
The above analysis turns out that
$\alpha(K_{\rho}(L))$ equals to zero exactly 
in the same way as for the case $U=0$\cite{chi}.
Therefore, $M(\chi_0H)=\chi_0 H-2(\chi_0H)^3+{\rm higher~orders~in}~H$ 
as described by a local Fermi liquid.

In conclusion, we have studied the ground state properties 
of a magnetic impurity
coupled to an interacting 1D system of conduction electrons.
 We found that a local Fermi liquid picture is still valid
but that the characteristic energy scale, i.e., $T_K$ is substantially 
enhanced by the strong repulsive 
interaction in the chain and affected non-monitonically in the case of electron attraction.
 We infer the validity of a
local Fermi liquid description from the fact that the magnetization
$M(\chi_0H)$ shows the same critical behavior for $U=0$ and for $U\neq 0$.

The author sincerely appreciates Prof. P. Fulde for 
instructive discussions, constant encouragement, 
kindly revising the manuscript and hospitality during his stay.
He is very grateful to Drs. H. Johannesson, T. Schork, S.-Q. Shen,
P. Thalmeier, D.F. Wang
and X. Zotos and Profs. D.L. Cox, A. Keller, I. Peschel,
P. Prelov\v sek, K.D. Schotte, H. Schulz, Z.B. Su and Lu Yu
for fruitful discussions and Dr. H. Scherrer for computational support. 
 
 \end{multicols}
 \begin{multicols}{2}
 \narrowtext
\begin{figure}
\caption {(a) $\chi_0\sim\xi_K$ and $\chi_0^{-1}\sim T_k$ {\it vs} 
$J$ at $U=0$. The left vertical axis is for  $\chi_0$ and the right one
for $\chi_0^{-1}$. (b) The same as (a) but at $U=20$.
The fit-curves are guides 
 to the eye (Solid curve and the filled circles are for $\chi_0$ 
and Dashed-line curve and the filled square for
$\chi_0^{-1}$).}
\label{fig1}
\caption { $\chi_0$ {\it vs} $U$ at $J\rho=2/\pi$.
The fit-curves are guides 
 to the eye.} 
\label{fig2}
\caption { Correlation functions between the impurity spin and
electron spins, and the local states between the impurity and the spin
at $i=0$ for $J=0.0005, 0.5$ at $U=20$.} 
\label{fig3}
\caption { $M(\chi_0H)$ {\it vs} $\chi_0H$.
The solid curve is for strong coupling limit. Each kind of
symbols for a given set of $J$ and $U$.
Inset: the amplified crossover regime.}
\label{fig4}
\end{figure}
 \end{multicols}

\begin{references}
\bibitem{Spin} As $L=4k+3$,
the number of sites in the two sides of the impurity differs
 by two sites, i.e., the systems are asymmetric, 
but the ground states still have $S^{total}=0$.
\bibitem{R199}At $U=0$, the precise scaling is achieved
only for $J, T, H\ll W$ and a flat density of states\cite{R13,TW,Hew}.
\bibitem{chi} 
We note here that the size-dependent $K_{\rho}(L)$ does not appear in the exponent
of $H$ in the small $H$ expansion {\it but} is non-trivially hidden 
in $\chi_0(L)$ which associates the screening cloud with $K_{\rho}(L)$.
\end{references}
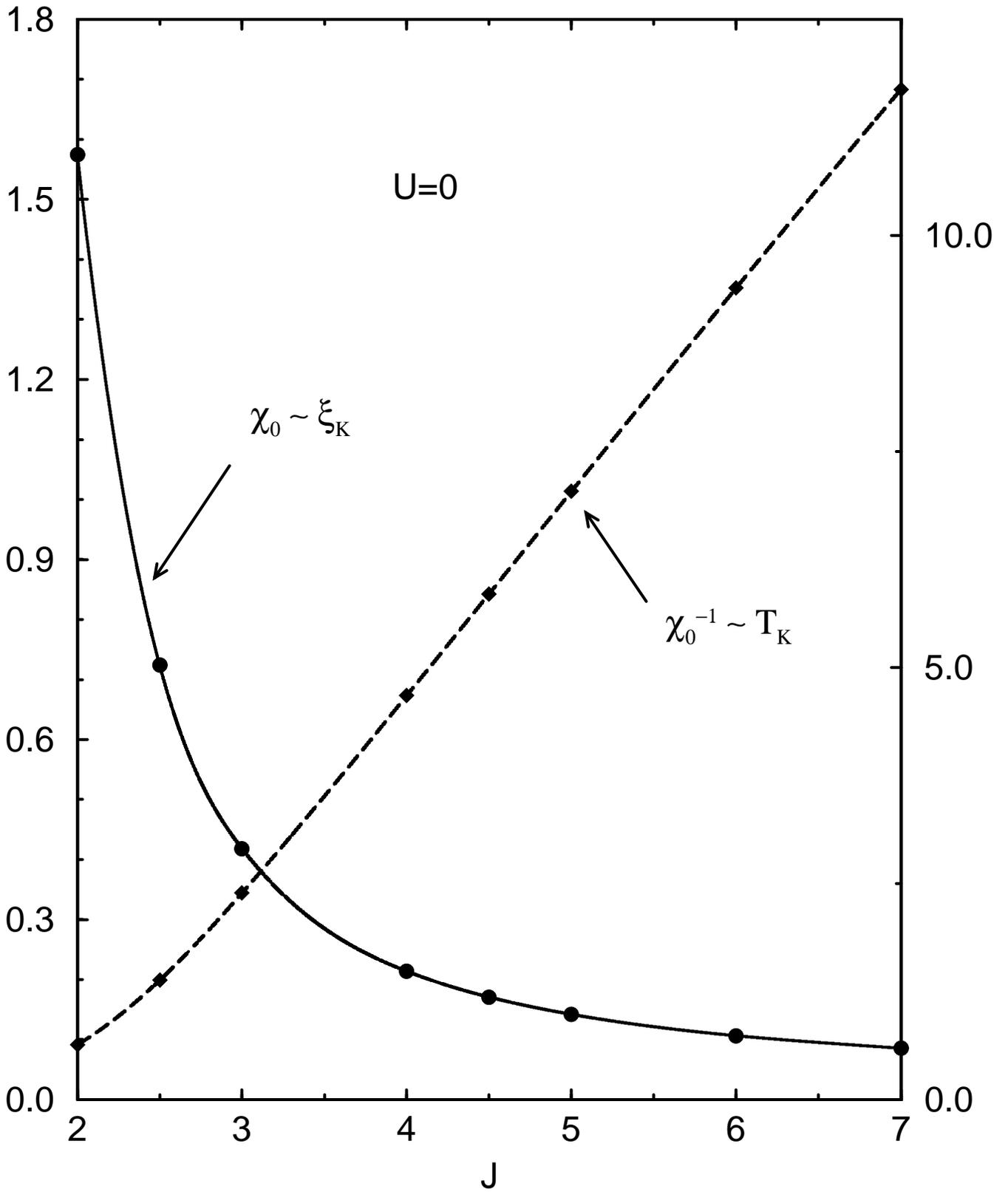
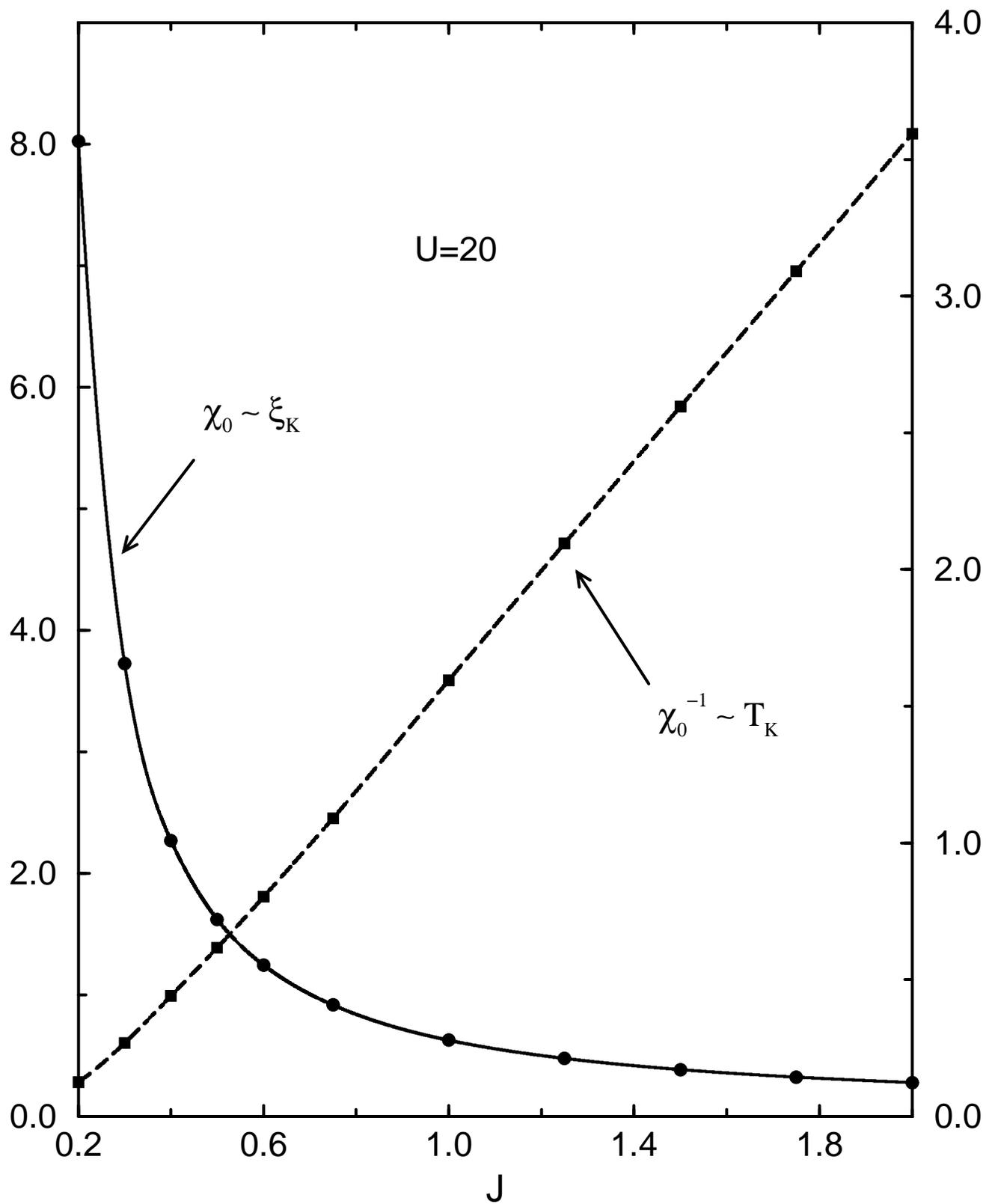
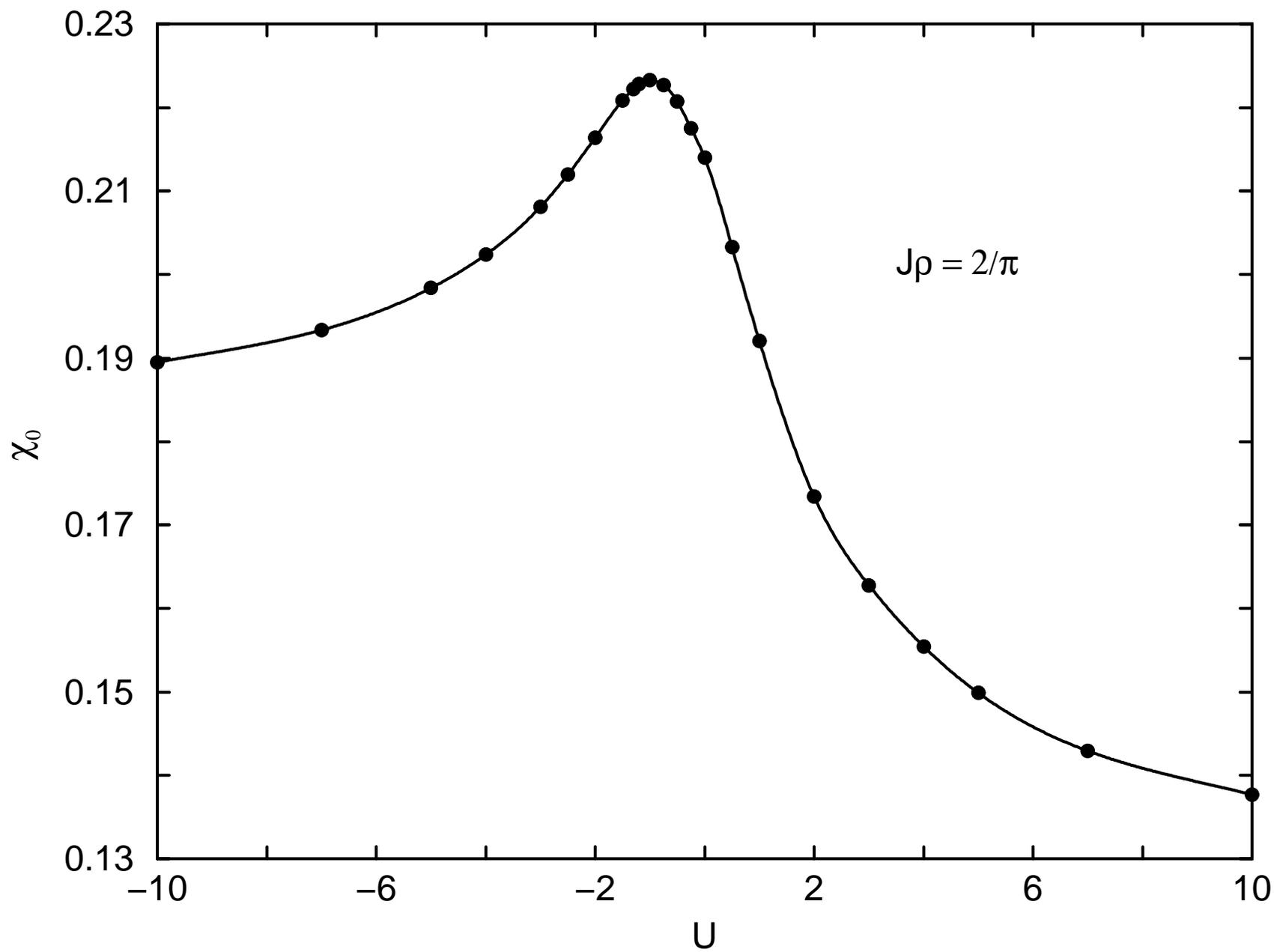
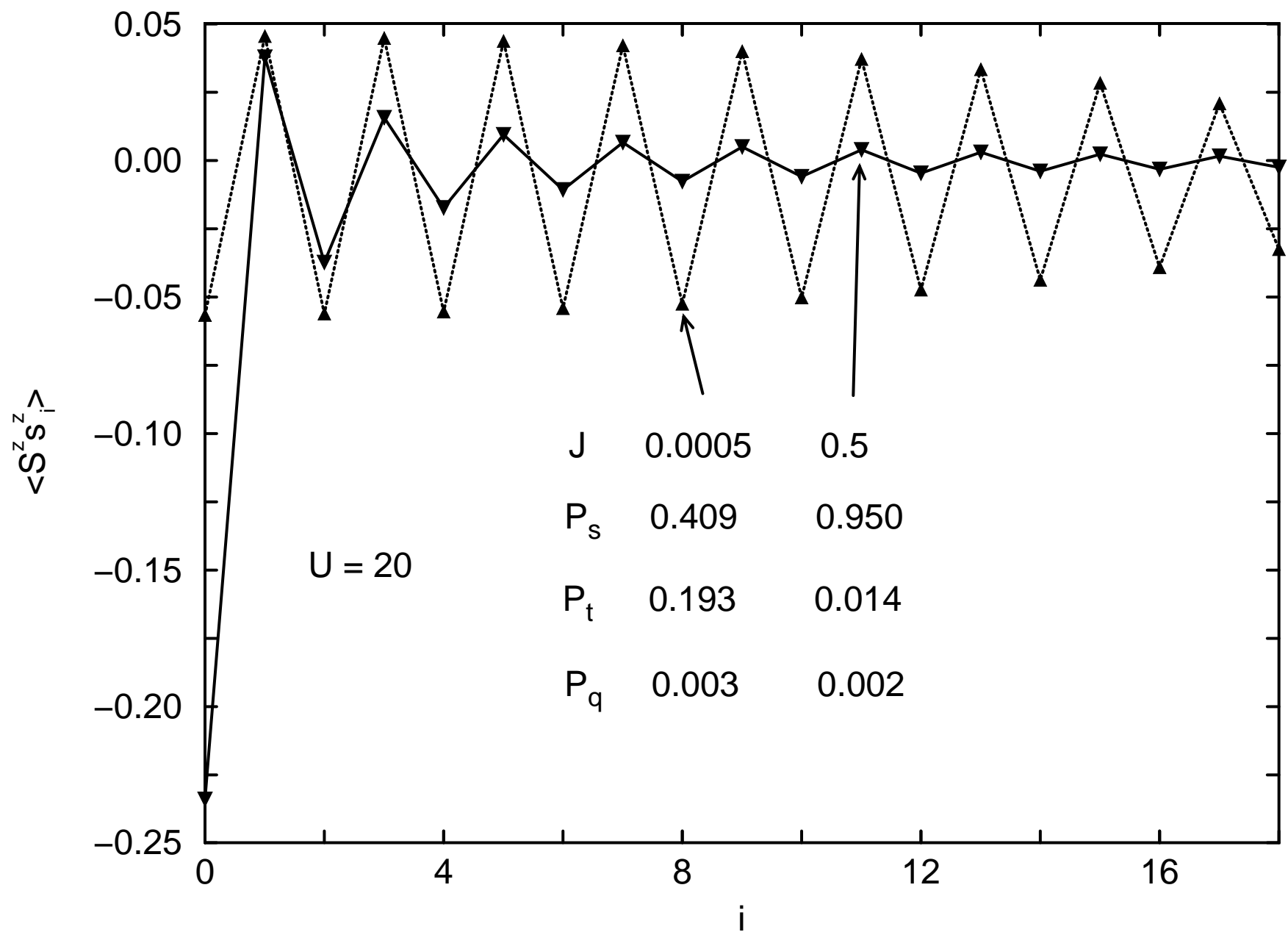
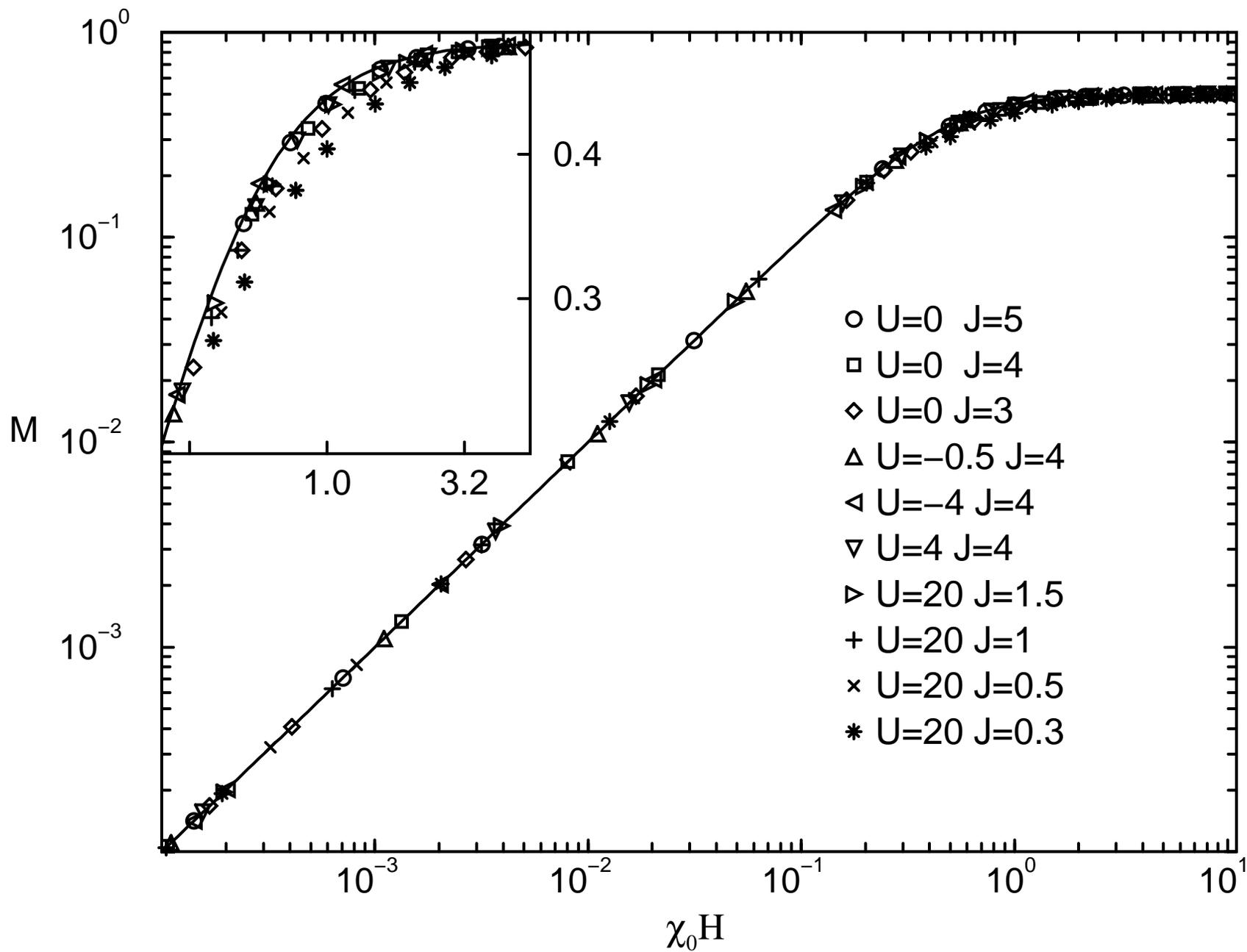
\end{document}